\documentclass[a4paper,twocolumn,11pt]{quantumarticle}
\pdfoutput=1
\usepackage{amssymb}
\usepackage{siunitx}
\usepackage{amsmath}
\usepackage{multirow} %
\usepackage{booktabs} 
\usepackage{bm} %
\newtheorem{myDef}{Definition}
\usepackage[numbers,sort&compress]{natbib}
\usepackage{makecell}
\usepackage{graphicx}%
\usepackage{bm}%
\usepackage{soul}
\usepackage[utf8]{inputenc}
\usepackage[english]{babel}
\usepackage[T1]{fontenc}
\usepackage{hyperref}
\usepackage{tikz}
\usepackage{lipsum}
\usepackage{hyphenat}

\makeatletter

\newcommand{\Rmnum}[1]{\expandafter\@slowromancap\romannumeral #1@}
\makeatother

\usepackage[ruled,linesnumbered]{algorithm2e}
\SetKwRepeat{Do}{do}{while}%
\begin{document}

\title{Appending Information Reconciliation for Quantum Key Distribution}

\author{Han Zhou}
\affiliation{%
College of Computer, National University of Defense Technology, Changsha 410073, China
}%

\author{Bang-Ying Tang}%
\affiliation{%
College of Computer, National University of Defense Technology, Changsha 410073, China
}%

\author{Huan Chen}
\affiliation{%
College of Liberal Arts and Sciences, National University of Defense Technology, Changsha 410073, China
}%

\author{Hui-Cun Yu}
\affiliation{%
Information and Navigation College, Air Force Engineering University, Xi'an 710077, China
}%
\affiliation{%
College of Advanced Interdisciplinary Studies, National University of Defense Technology, Changsha 410073, China
}%
\author{Si-Chen Li}  
\affiliation{%
College of Computer, National University of Defense Technology, Changsha 410073, China
}%
\author{Wan-Rong Yu} 
\affiliation{%
College of Computer, National University of Defense Technology, Changsha 410073, China
}%
\email{wlyu@nudt.edu.cn}
\author{Bo Liu}
\affiliation{%
College of Advanced Interdisciplinary Studies, National University of Defense Technology, Changsha 410073, China
}%
\email{liubo08@nudt.edu.cn}

\maketitle

\begin{abstract}
	Information reconciliation (IR), which corrects the errors in the sifted keys, directly determines the secure key rate and the link distance of quantum key distribution (QKD) systems. In this article, we propose an appending information reconciliation (AIR) scheme based on polar codes, which achieves high efficiency and ultra-low failure probability simultaneously, by gradually disclosing the bit values of the polarized channels with high error probability. The experimental results show that the efficiency of the proposed AIR scheme is closer to the Shannon limit, compared with the state-of-the-art implemented polar codes-based IR schemes, with the overall failure probability around $10^{-8}$, especially when performed with smaller block sizes. Moreover, the efficiency of the proposed AIR scheme is 1.046, when the block size is \SI{1}{Gb} and the quantum bit error rate of 0.02. Therefore, the proposed AIR scheme can further eradicate the performance gap between theory and implementation for QKD systems.
 	
\end{abstract}

\section{Introduction}

Quantum key distribution (QKD), based on the laws of quantum physics, can distribute the information-theoretical-secure keys between Alice and Bob, even in the presence of an eavesdropper with unlimited computational power~\cite{RN18,RN96}. Information reconciliation (IR) is a key post-processing procedure of practical QKD systems, which corrects the errors bits in sifted keys, caused by imperfect implementations and  attacks~\cite{RN70}.

In the IR procedure, the sifted keys ($K^A_s$ and $K^B_s$) are reconciled to the equal and weak secure keys $K_\mathrm{IR}$, by exchanging the syndrome information via a classical public channel~\cite{RN1}. According to the way of syndrome information interaction, IR schemes can be divided into interactive IR schemes and one-way IR schemes~\cite{RN30,RN97}. The interactive IR schemes (Cascade) can reach the efficiency of $1.03$, however, the heavy latency of classical communication has to be suffered with around $14$ interactive rounds~\cite{RN7,RN27,RN23,RN60}. The one-way IR schemes are based on the forward error correction codes, such as low-density parity-check codes (LDPC)~\cite{RN14,RN94,RN99} and polar codes~\cite{RN4,RN12,RN23} \textit{etc}, which significantly decrease the communication latency by exchanging only one syndrome. Especially, polar codes-based IR schemes, with the potential to approach the Shannon-limit efficiency, have been adapted into QKD systems to further improve the secure key rate~\cite{RN12}. 

The first polar codes-based IR scheme is implemented by P. Jouguet and S. Kunz-Jacques in 2014, which reached the efficiency of $1.12$ with the failure probability $\varepsilon = 0.1$ and the block size of~\SI{16}{Mb}~\cite{RN12}. However, the application of polar codes-based IR schemes for QKD systems is limited by the high failure probability~\cite{RN65,RN25}. In 2018, Yan \textit{et al.} improved the polar codes-based IR scheme with the successive cancellation list decoding strategy, which decreased the $\varepsilon$ to $10^{-3}$ with the block size of \SI{1}{Mb}~\cite{RN25}. On the other hand, in 2021, Tang \textit{et al}. proposed the Shannon-Limit approached polar codes-based scheme by introducing an acknowledgment reconciliation phase, which decreased the $\varepsilon$ to $10^{-8}$ and the efficiency is improved to 1.091 with the IR block size
of 128Mb~\cite{RN2}. However, the success of Tang's acknowledgment reconciliation procedure was guaranteed by the LDPC codes, which can not take full advantage of the information laid in the polarized channels.

In this article, we propose a novel appending IR (AIR) scheme based on the polar codes, which achieves high efficiency and ultra-low failure probability simultaneously, by gradually disclosing the bit values of the polarized channels with high error probability. First of all, for given quantum bit error rates and the target $\varepsilon$, the frozen vectors can be optimized and pre-shared between Alice and Bob. Then, the certain bit values of the polarized channels, which are suffering high error probability, will be combined as the syndrome by Alice and transmitted to Bob. Afterwards, Bob decides to request for appending more syndrome information from Alice, or to abort the IR procedure once comes up to the maximum interactive rounds, by comparing the cyclic redundancy check values between the decoded and Alice's encoded codewords.

The experimental results show that the efficiency of the proposed AIR scheme is further closer to the Shannon limit, compared with the state-of-the-art implemented polar codes-based IR schemes, with the overall failure probability around $10^{-8}$, especially when performed with smaller block sizes. Moreover, the efficiency of the proposed AIR scheme is 1.046, when the block size is \SI{1}{Gb} and the quantum bit error rate of 0.02. Thus, our proposed AIR scheme can further eradicate the performance gap between theory and implementation for QKD systems.

\section{Preliminaries}
\subsection{Information Reconciliation}

In quantum key distribution (QKD) systems, information reconciliation (IR) corrects the errors in the sifted keys into an equal and weak secure reconciled key, by exchanging the syndrome information through the classical channel~\cite{RN153, RN2}.

Assume Alice's (Bob's) sifted key is $K_s^A$ ($K_s^B$) with the length of $n$, the reconciled key is $K_\mathrm{IR}^A$ and $K_\mathrm{IR}^B$.

The performance of IR schemes for QKD systems is mainly evaluated by the efficiency $f$ and the failure probability $\varepsilon$.

Assume the amount of the leaked key information to eavesdroppers is $m$, the efficiency $f$ is defined as
\begin{equation}\label{eq:f}
f(E_{\mu})=\frac{m}{nH_2(E_{\mu})},
\end{equation}
where $E_\mu$ is the quantum bit error rate (QBER) and $H_2(x)$ is the binary Shannon entropy, which can calculated as
\begin{equation}\label{eq:h2}
H_2(x)=-x\log_2(x)-(1-x)\log_2(1-x).
\end{equation}

After the IR procedure, there is a small failure probability that $K_\mathrm{IR}^A \neq K_\mathrm{IR}^B$
\begin{equation}\label{eq1}
\mathrm{Pr}(K^A_\mathrm{IR} \neq K^B_\mathrm{IR}) \leq \varepsilon.
\end{equation}

\subsection{Polar Codes-based IR Scheme}

In 2009, E. Arikan proposed the polar codes, which could achieve the symmetric capacity of any given binary-input discrete memoryless channel (BDMC) by polarizing the channels~\cite{RN4}. P. Jouguet and K.-J. Sebastien first performed polar codes-based IR scheme for QKD systems in 2014~\cite{RN12}.

In polar codes, individual $n$ copies of BDMCs are polarized to the $q$ noisy channels (frozen bits) and $n-q$ error-free channels (information bits). The locations of the frozen bits, defined as the frozen vector $V$, can be determined by selecting the $q$ channels with the high maximum likelihood decoding error probability~\cite{RN37}.

Firstly, Alice encodes the sifted key $K_s^A$ into the codeword $U$ by
\begin{equation}
\label{equ:encode}
U=K_s^A G^n=K_s^A F^{\otimes \log n}B_n,
\end{equation}
where $G^n$ is the polar transform matrix, $F = \begin{bmatrix} 1 & 0\\1 & 1 \end{bmatrix}$ and $B_n$ is the permutation matrix for bit-reversal operation~\cite{RN4}.

\begin{figure*}[!htb]
	\centering
	\includegraphics[width=0.9\textwidth]{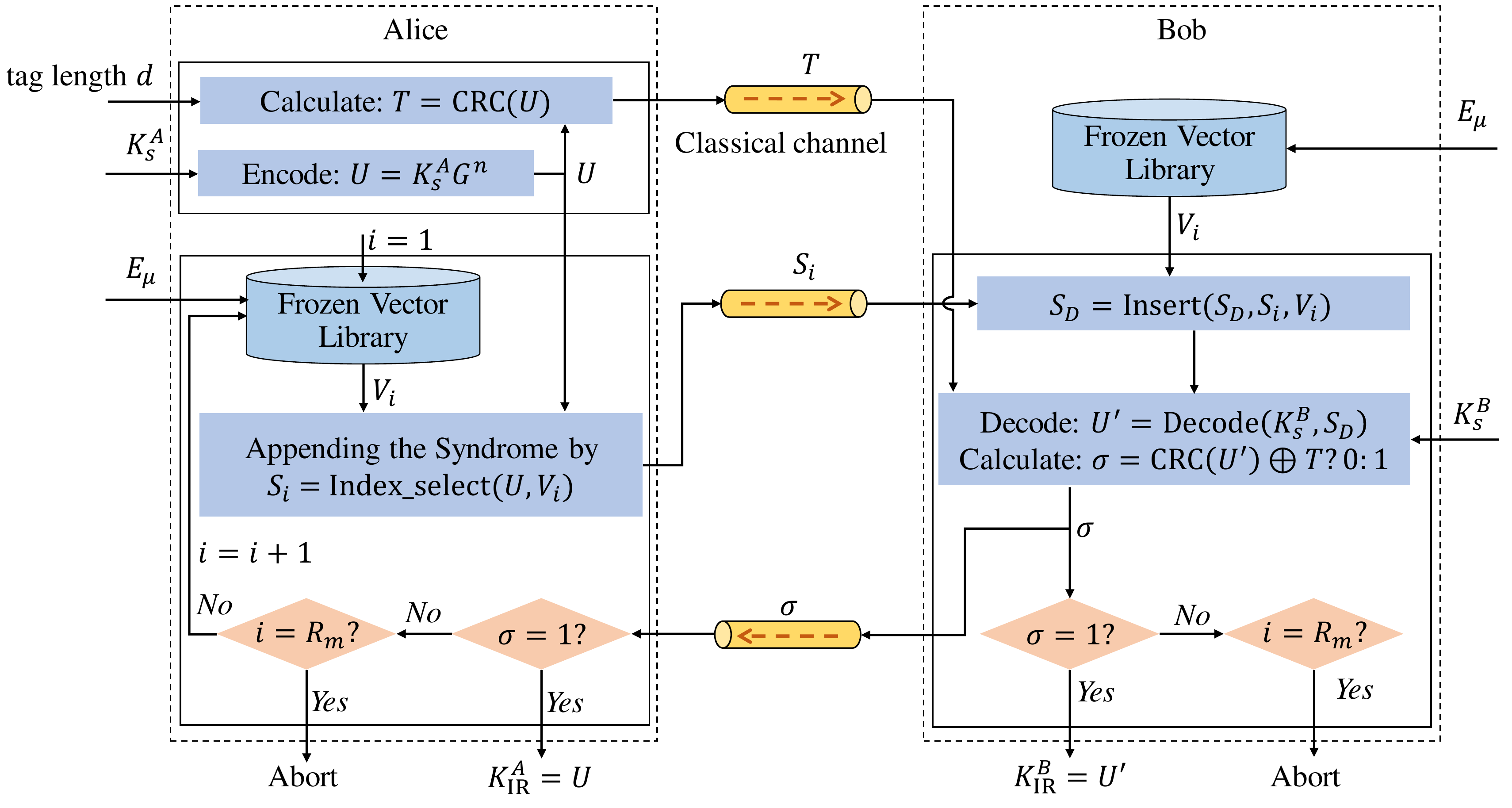}
	\caption{The schematic diagram of the proposed AIR scheme. $G^n$: the polar transform matrix, $K_s^A$ and $K_s^B$: the sifted key of Alice and Bob, $E_\mu$: quantum bit error rate, CRC: cyclic redundancy check, $K_\mathrm{IR}^A$ and $K_\mathrm{IR}^B$: the reconciled secure key of Alice and Bob.
	}
	\label{fig1}
\end{figure*}

Then, the syndrome $S$ is combined with the frozen bits of the codeword $U$, indicated by $V$.

Afterwards, Bob corrects the errors in $K_s^B$ with $S$. This procedure is also called decoding. The first proposed decoder of polar codes, named successive cancellation (SC) decoder, recursively calculates the likelihood probability of information bits and makes a hard decision for the outputs~\cite{RN4}.

However, SC decoding is a greedy tree search algorithm which can only achieve a local optimum, resulting in high failure probability. Afterwards, the successive cancellation list (SCL) highly improves the performance of polar codes, by decoding $L$ temporal vectors and selecting the one with the highest maximum likelihood decoding probability as the decoded vector.

Furthermore, the upper bound of the decoding failure probability could be estimated by~\cite{RN37}
\begin{equation}
  \label{eq:varepsilon}
  \varepsilon \leq \sum _{i \notin V} P_e^i,
\end{equation}
where $P_e ^ i$ is the error probability under maximum-likelihood decision of the $i$-th information bit, $i \in \{0,1,\cdots,n-1\}$.

\section{The Proposed Appending Information Reconciliation Scheme}

In this article, we propose an appending IR (AIR) scheme based on polar codes, and the schematic diagram is shown in FIG.~\ref{fig1}. First of all, the frozen vector library for given different $n$ and $E_\mu$ should be pre-shared between Alice and Bob. The optimization of frozen vectors for given overall failure probability $\varepsilon$ is detailed described in section~\ref{optimized_V}. 

The implementation of the proposed AIR scheme mainly includes two phases: initialization phase and appending reconciliation phase. In the initialization phase, Alice encodes the sifted key $K_s^A$ into the codeword $U$ and calculates the CRC tag $T$ of $U$. Afterwards, the CRC tag $T$ is transmitted to Bob. For each round $i$ of the appending reconciliation phase, Alice appends the syndrome $S_i$ with the optimized frozen vector $V_i$. After received $S_i$, Bob decodes the sifted key $K_s^B$ into $U^\prime$ and calculates the CRC tag of $U^\prime$. Then Bob decides to request for appending more syndrome information or to abort the IR procedure once comes up to the maximum interactive rounds.

\subsection{Initialization}

Alice encodes the sifted key $K_s^A$ into the codeword $U$ by Eq.~(\ref{equ:encode}). Then, for given tag length $d$, the CRC tag $T$ of $U$ is calculated and transmitted to Bob via the classical channel. 

\subsection{Appending reconciliation}

For each round $i$, $1\leq i \leq R_m$, of the appending reconciliation phase, Alice and Bob perform the following steps, where $R_m$ is the maximum interactive round number.   

Step 1, Alice appends the syndrome $S_i$ by picking up the bit values from $U$ with the optimized frozen vector $V_i$, described as $S_i = \mathrm{Index\_ select}(U, V_i)$.

Step 2, Alice sends $S_i$ to Bob via the classical channel.

Step 3, Bob updates the integrated syndrome string $S_D$, by replacing the bit values, indicated by the frozen vector $V_i$, with the corresponding value of $S_i$. This step is described as $S_D=\mathrm{Insert}(S_D,S_i,V_i)$. Here $S_D$ is an $n$-length vector, where the elements representing frozen bits are set to ``$0$'' or ``$1$'' and the rest are set to ``$-1$''. $S_D$ is initialized to $\{-1\}^n$. 

Step 4, Bob decodes $K^B_s$ into $U^\prime$ with $S_D$, and calculates the flag value by $\sigma = (\mathrm{CRC}(U^\prime)\oplus T)?0:1$, where $\sigma = 1$ when the CRC tag of the decoded vector $U^\prime$ is equal to $T$, otherwise $\sigma = 0$. The decoding procedure of $K_s^B$ can be described as $ U^\prime= \mathrm{Decode}(K_s^B,S_D)$. In this article, we work with the SCL decoder.

Step 5, Bob returns the flag value $\sigma$ to Alice. 

Step 6, If $\sigma=1$, the IR procedure will end by outputting the reconciled keys, where $K_\mathrm{IR}^A = U$ and $K_\mathrm{IR}^B = U^\prime$. If $\sigma=0$ and $i=R_m$, the IR procedure is failed. Otherwise, Alice and Bob start the next appending reconciliation round and back to Step 1.

\subsection{Performance Analysis of the AIR Scheme}

Assume that $\varepsilon_i$ is the decoding failure probability of the SC or SCL decoder with the $S_D^i$, which is the syndrome string $S_D$ used in the $i$-th round.

The decoding success probability in the $i$-th round, which means the decoding procedure fails with the $S_D^{i-1}$ and succeeds with the $S_D^{i}$, can be calculated as
\begin{equation}
\begin{aligned}
\mathrm{Pr}[(U^\prime = U | S_D^i) \cap (U^\prime \neq U | S_D^{i-1})] = \varepsilon_{i-1} - \varepsilon_i
\end{aligned},
\end{equation}
due to $S_D^i$ is appended more frozen bits to $S_D^{i-1}$, where $1 \leq i \leq R_m$ and $\varepsilon_0=1$. The detailed analysis is given in Appendix~\ref{sectionAP:success probability}.

For the codeword $U$ and the decoded codeword $U^\prime$, the CRC failure probability in our proposed AIR scheme can be defined as
\begin{equation}
  \label{eq:pcrc}
  \begin{aligned}
    \varepsilon_{\mathrm{CRC}} &= \mathrm{Pr}\left[\mathrm{CRC}(U^\prime)
    =\mathrm{CRC}(U)|U^\prime \neq U\right]\\ 
    &\leq \frac{L}{2^d}\\
  \end{aligned},
\end{equation}
where $L$ is the list size of SCL decoder and $d$ is the length of the CRC tag. For the SC decoder, $L=1$.

\subsubsection{The overall failure probability}

Assume that the AIR scheme ends in the $i$-th round, the overall failure probability of the AIR scheme can be analyzed in two cases.

\textbf{Case \Rmnum{1}.} $\mathrm{CRC}(U^{\prime})=T$ but $U^\prime \neq U$ and $i\leq R_m$, which means the IR procedure fails the CRC check in the previous $(i-1)$ rounds and succeeds in the $i$-th round, but the decoding procedure fails. The failure probability $\varepsilon_\mathrm{I}$ of this case is calculated as

\begin{equation} 
  \label{eq:pf1}
\begin{aligned}
  \varepsilon_\mathrm{I} = \varepsilon_\mathrm{CRC}\sum_{i=1}^{R_m}{\varepsilon_i(1-\varepsilon_\mathrm{CRC})^{i-1}}
\end{aligned}.  
\end{equation}

\textbf{Case \Rmnum{2}.} $\mathrm{CRC(U^{\prime})}\neq T$ and $i = R_m$, which means the IR procedure fails the CRC cheack in the previous ($i-1$) rounds and aborts in the maximum round $R_m$. In this case, the failure probability $\varepsilon_\mathrm{II}$ can be calculated as
\begin{equation}
  \label{eq:pf2}
  \begin{aligned}
  \varepsilon_\mathrm{II} = \varepsilon_{i}(1-\varepsilon_\mathrm{CRC})^{i}
  \end{aligned}.  
\end{equation}

The detailed calculation of Eq.~(\ref{eq:pf1}) and Eq.(\ref{eq:pf2}) can be found in Appendix~\ref{sectionAP:pf}.

Therefore, the overall failure probability of the AIR scheme can be calculated as
\begin{equation}
  \label{eq:pf}
  \begin{aligned}
    \varepsilon &= \varepsilon_\mathrm{I} + \varepsilon_\mathrm{II}\\
    &=\varepsilon_\mathrm{CRC}\sum_{i=1}^{R_m}{\varepsilon_i(1-\varepsilon_\mathrm{CRC})^{i-1}} + \varepsilon_{R_m}(1-\varepsilon_\mathrm{CRC})^{R_m}\\
    &<R_m\varepsilon_\mathrm{CRC}+\varepsilon_{R_m}\\
  \end{aligned}.
\end{equation}

When the length of the CRC tag $d$ is large enough, $R_m\varepsilon_\mathrm{CRC} \ll \varepsilon_{R_m}$, which will result in $\varepsilon \lesssim \varepsilon_{R_m}$.

\subsubsection{The average execution rounds}\label{section:round_av}

Assume the AIR scheme is stopped in $i$-th round with probability of $P_s^i$, where $1\leq i \leq R_m$. Then, $P_s^i$ can be analyzed in two cases.

\textbf{Case \Rmnum{1}.} $i<R_m$, which means the IR procedure fails the CRC check in the previous $(i-1)$ rounds and succeeds in the $i$-th round. The $P_s^i$ can be calculated as
\begin{equation}
  \label{eq:round_av1}
  \begin{aligned}
  P_s^i =\varepsilon_{i-1}(1-\varepsilon_{\mathrm{CRC}})^{i-1} - \varepsilon_ {i}(1-\varepsilon_{\mathrm{CRC}})^{i}   
  \end{aligned}.
\end{equation}

\textbf{Case \Rmnum{2}.} $i=R_m$, which means the IR procedure is stopped in the the $R_m$-th round, the $P_s^i$ can be calculated as
\begin{equation}
  \label{eq:round_av2}
  \begin{aligned}
    P_s^i =\varepsilon_{i-1}(1-\varepsilon_{\mathrm{CRC}})^{i-1}\\
  \end{aligned}.
\end{equation}

The detailed analysis of Eq.~(\ref{eq:round_av1}) and Eq.(\ref{eq:round_av2}) is shown in Appendix~\ref{sectionAP:round_av}.

Therefore, the average execution round number can be calculated as
\begin{equation}
  \label{eq:R_av}
  \begin{aligned}
  \overline{R} &=\sum _{i=1} ^{R_{m}}iP_s^i \\
   &= \sum _{i=1} ^{R_{m} }i[\varepsilon_{i-1}(1-\varepsilon_{\mathrm{CRC}})^{i-1}]\\
   &-\sum _{i=1} ^{R_{m}-1}i[\varepsilon_{i}(1-\varepsilon_{\mathrm{CRC}})^{i}]
  \end{aligned}.
\end{equation}

\subsubsection{The reconciliation efficiency}
The average leakage information $m_i$ in $i$-th round could be calculated as 
\begin{equation}
  \label{eq:mi}
    m_i =q_i P_s^i,
  \end{equation}
where $q_i$ is the size of frozen vector used in $i$-th round.

According to the Eq.~(\ref{eq:f}), the efficiency of the AIR scheme can be expressed as 
\begin{equation}
  \label{eq:f_air}
  \begin{aligned}
    f&=\frac{d+\sum _{i=1} ^{R_{m}} m_i}{nH_2(E_\mu)}\\
    &=\frac{d+\sum _{i=1} ^{R_{m}} q_i \varepsilon_{i-1}(1-\varepsilon_\mathrm{CRC})^{i-1}}{nH_2(E_\mu)}\\
    &-\frac{\sum _{i=1} ^{R_{m}-1} q_i \varepsilon_{i}(1-\varepsilon_\mathrm{CRC})^{i}}{nH_2(E_\mu)}.\\
  \end{aligned}
\end{equation}

\subsection{Optimization of frozen vectors}
\label{optimized_V}

Given the block size $n$ and the QBER $E_\mu$, the upper bound of error probability $P_e^i$ in $i$-th polarized channel can be calculated by the degrading and upgrading quantizations~\cite{RN37}.

According to Eq.~(\ref{eq:varepsilon}), the decoding failure probability will be decreased with the larger size of the frozen bits. Thus, given the maximum interactive round number $R_m$, the frozen vector used in each round can be optimized.

\begin{figure}[!ht]
  \centering
  \includegraphics[width=0.48\textwidth]{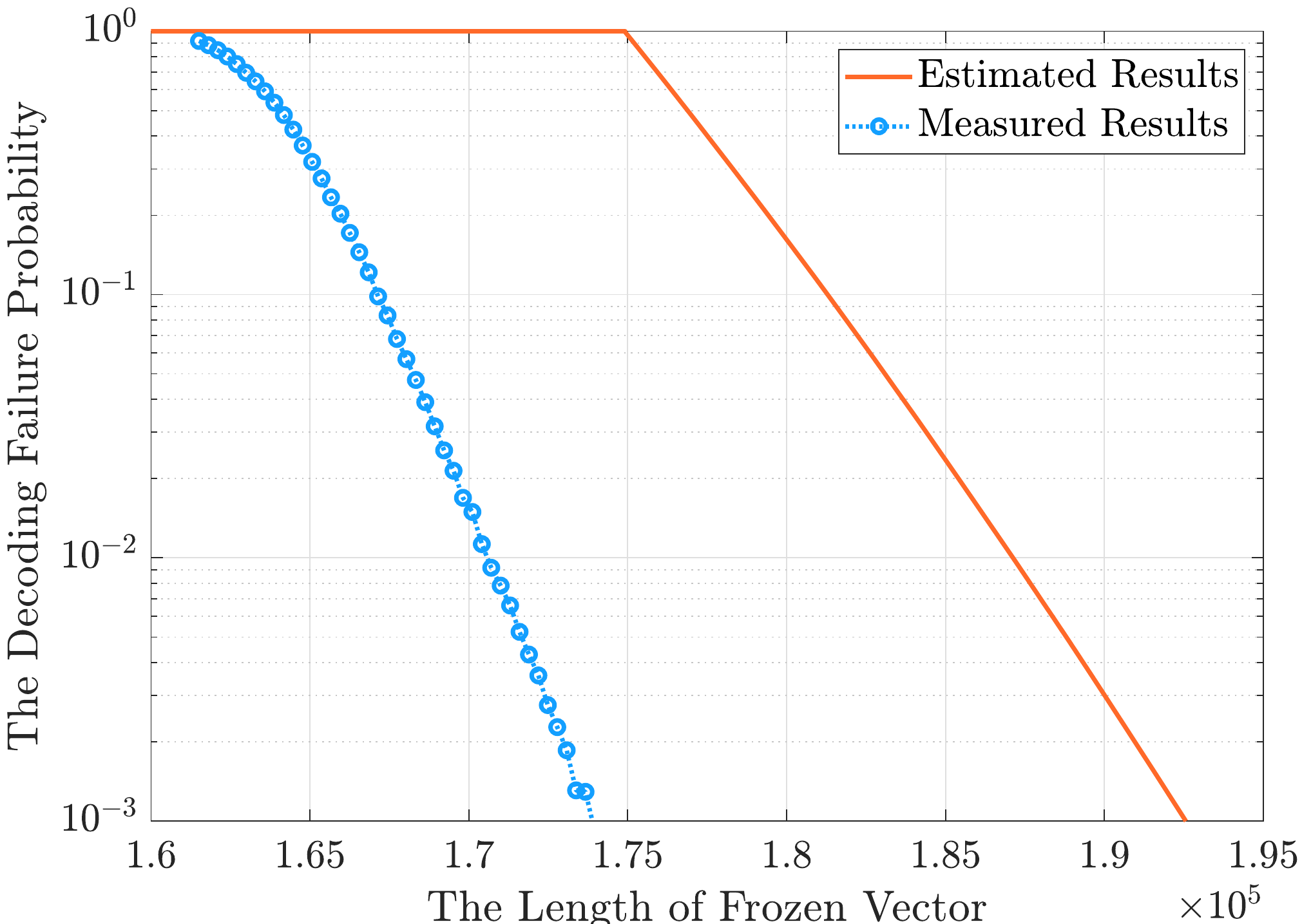}
  \caption{The estimated and measured decoding failure probability with different length of frozen vectors. The block size is \SI{1}{Mb}, $E_\mu = 0.02$, the length of frozen vector is varying from $1.60\times 10^5$ to $1.95\times 10^5$, with step size $\beta = 300$. The list size of SCL decoder $L$ is set as 16.}
  \label{fig.K_fer}
  \end{figure}

We show the estimated and the measured decoding failure probability with given frozen vector according to Eq.~(\ref{eq:varepsilon}) in FIG.~\ref{fig.K_fer}, where the block size is \SI{1}{Mb}, $E_\mu=0.02$, $L=16$. The length of frozen vector is varying from $1.60\times 10^5$ to $1.95\times 10^5$, with step size $\beta = 300$. The measured decoding failure probability is rapidly decreased to $10^{-3}$, while the estimated $\varepsilon=1.0$.

\begin{myDef}\label{def_op1}
  $(P_e,W) = {\rm Descend}(P_e)$ is defined as sorting the error probability vector $P_e$ in descending order and recording the corresponding locations into the vector $W$.
  \end{myDef}

\begin{myDef}\label{def_op5}
  Assume that $\mathcal{B}=\left\{(q,\varepsilon)|q\in \mathbb{N}, \varepsilon\in[0,1]\right\}$, where $q$ is the frozen vector size and $\varepsilon$ is the corresponding decoding failure probability. $B=\mathcal{M}(\mathcal{B},\delta)$ represents that $B \subseteq  \mathcal{B}$ and $B=\{(q_i,\varepsilon_i)|(q_i,\varepsilon_i) \in \mathcal{B}, q_{1}<q_2<\cdots < q_\delta,\delta\leq |\mathcal{B}|\}$.
  \end{myDef}

\begin{myDef}
$(K_s^A, K_s^B)= {\rm{Rand}}(n,E_\mu)$ is defined as randomly generating key string $K_s^A$ and $K_s^B$ with length of $n$ and QBER of $E_\mu$.
\end{myDef}

\begin{algorithm*}
  \caption{Optimized Frozen Vectors Procedure} 
  \label{AL1}  
  \KwIn{$n$, $E_\mu$, $R_m$, $\varepsilon_{R_m}$,$P_e$,  $\beta$ and the round number $t$ for measuring the decoding failure probability} 
  \KwOut{the frozen vectors $V_1$, $V_2$,$\cdots$ and $V_{R_m}$} 
  $(P_e,W) = {\rm Descend}(P_e)$, $p=1$, $\mathcal{B}=\varnothing$, $\alpha = \lceil nH_2(E_\mu)\rceil$\;
  \While{$p \neq 0$}
  {
    $\mathcal{B} \leftarrow \mathcal{B}\cup (\alpha,p)$, $\alpha \leftarrow \alpha + \beta$, $S_D=\{-1\}^n$, $e_{\rm{cnt}}=0$\;
    \For{$i=1$ to $t$}
    {
	    $(K_s^A,K_s^B) = {\rm{Rand}}(n,E_\mu)$, $U=K_s^AG^n$\;
	    $S=\mathrm{Index\_ select}(U,W[0:\alpha])$, $S_D=\mathrm{Insert}(S_D,S,W[0:\alpha])$\;
	    \textbf{if} ${\rm Decode}(K_s^B,S_D) \neq U$ \textbf{then} $e_{\rm{cnt}} \leftarrow e_{\rm{cnt}}+1$\;
    }
    $p=e_{\rm{cnt}}/t$\;
  }
  $q =n - \max\{i|\sum_{j=0}^{i} P_e[n-j]\leq \varepsilon_{R_m}, i\in(0,n-\alpha)\}$, $B_{\mathrm{opt}} = \mathrm{Opti\_Effi}(\mathcal{B},R_m,q,\varepsilon_{R_m})$\;
  $V_1 = W[0:q_1]$, $V_i=W[q_{i-1},q_i]$, where $(q_i,\varepsilon_i) \in B_{\mathrm{opt}}$ and $ 2 \leq i \leq R_m$\; 
  \end{algorithm*}

\begin{myDef}
	$B_{\mathrm{opt}} = \mathrm{Opti\_Effi}(\mathcal{B}, R_m,q,\varepsilon_{R_m})$ is defined as finding out the optimal $B$, $B=\mathcal{M}(\mathcal{B},R_m -1) \cup (q,\varepsilon_{R_m})$, which resulting the efficiency $f$ closest to the Shannon-limit with Eq. (\ref{eq:f_air}).
\end{myDef}

The detailed description of the optimized procedure of frozen vectors is shown in Algorithm~\ref{AL1}.

  \section{Results}

  We have implemented the proposed AIR scheme with the detailed parameters shown in Table~\ref{tab.1M parameters}. 
  
    \begin{table}[htbp]
  	\begin{center}
  		\setlength{\tabcolsep}{5pt}{%
  		\renewcommand{\arraystretch}{1.2} %
  		\caption{The parameters for the proposed AIR scheme.} 
  		\label{tab.1M parameters}
  		\begin{tabular}{c c} 
  			\toprule[1pt]
  			$n$ & $2^{16} \sim 2^{30}$ \\
  			\hline
  			$E_\mu$ & $0.01 \sim 0.12$\\
  			\hline
  			$R_m$ & $2 \sim 6$\\
  			\hline
  			$\varepsilon_{R_m}$ & $10^{-8}$\\
  			\hline
  			$d$ & 64\\\hline
  			Decoder & SCL\\\hline
  			$L$ & $16$\\
  			\hline
  			$V$ & Optimized with Algorithm~\ref{AL1}\\\hline
  		\end{tabular}}
  	\end{center}
  \end{table}
  
\subsection{Efficiency with different maximum rounds\label{section:f_Rm}}

\begin{figure}[!ht]
  \centering
  \includegraphics[width=0.48\textwidth]{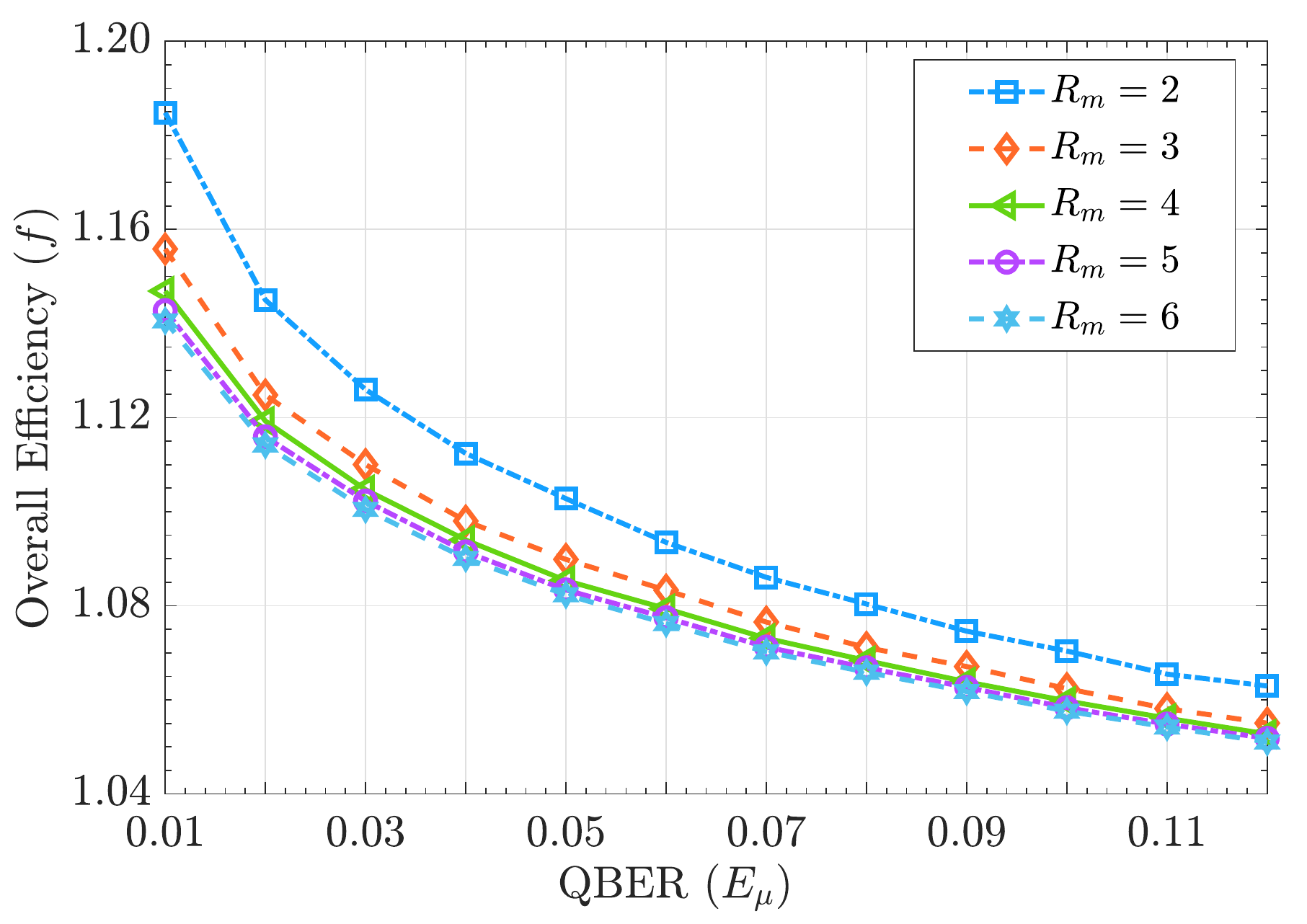}
  \caption{The efficiency of the AIR scheme with QBER from 0.01 to 0.12 and the block size of \SI{1}{Mb}, when the maximum round $R_{m} = 2,3,\dots,6$.}
  \label{fig.round_f}
  \end{figure}

For the certain block size $n$ and the QBER $E_\mu$, the efficiency of the AIR scheme is varying a lot when given different $R_m$, as shown in FIG.~\ref{fig.round_f}, where $n=$\SI{1}{Mb} $E_\mu=0.02$ and $R_{m} = 2,3,\dots,6$.

The appended length of frozen vector in each round will be decreased with larger $R_m$, which results in leaking less extra key information to eavesdroppers when correcting errors in the sifted keys. As shown in FIG.~\ref{fig.round_f}, the efficiency of the proposed AIR scheme is improved significantly when $R_m$ is increased from 2 to 4. When $R_{m} > 4$, although the efficiency can further slightly approach the Shannon limit, heavy communication latency would be caused by amounts of interactive rounds.

\subsection{Average execution round number}

\begin{figure}[!ht]
  \centering
  \includegraphics[width=0.48\textwidth]{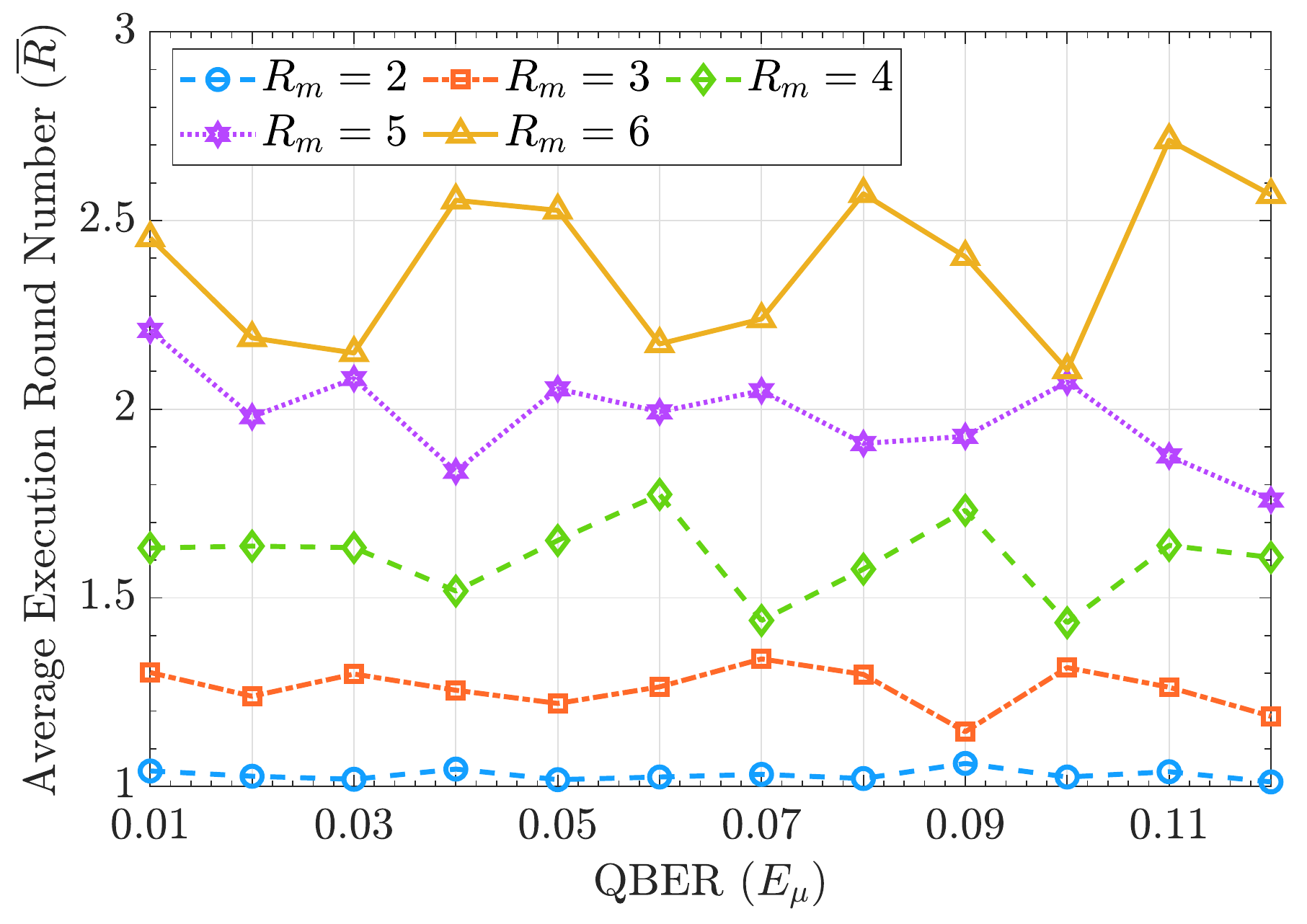}
  \caption{The average execution round number of the AIR scheme with $n=$\SI{1}{Mb}, $R_{m} = 2,3,\dots,6$ and $E_\mu$ is varying from 0.01 to 0.12.}
  \label{fig.round_av}
\end{figure}

Given the parameters in Table~\ref{tab.1M parameters}, the average execution round number $\overline{R}$ can be estimated by Eq.~(\ref{eq:R_av}). Here, the estimated $\overline{R}$ is shown in FIG.~\ref{fig.round_av} with $n=$\SI{1}{Mb}, $R_{m} = 2,3,\dots,6$ and $E_\mu$ is varying from 0.01 to 0.12. The average execution round number is less than 2 when $R_m \leq 4$. Thus, we suggest that $R_m\leq 4$ when implementing the proposed AIR scheme, to achieve high efficiency and less communication latency simultaneously. More detailed analysis results about $\overline{R}$ can be found in Appendix \ref{sec:appA}.

\subsection{Comparison with the previous schemes}
\label{section:opvious}

\begin{figure}[!ht]
	\centering
	\includegraphics[width=0.48\textwidth]{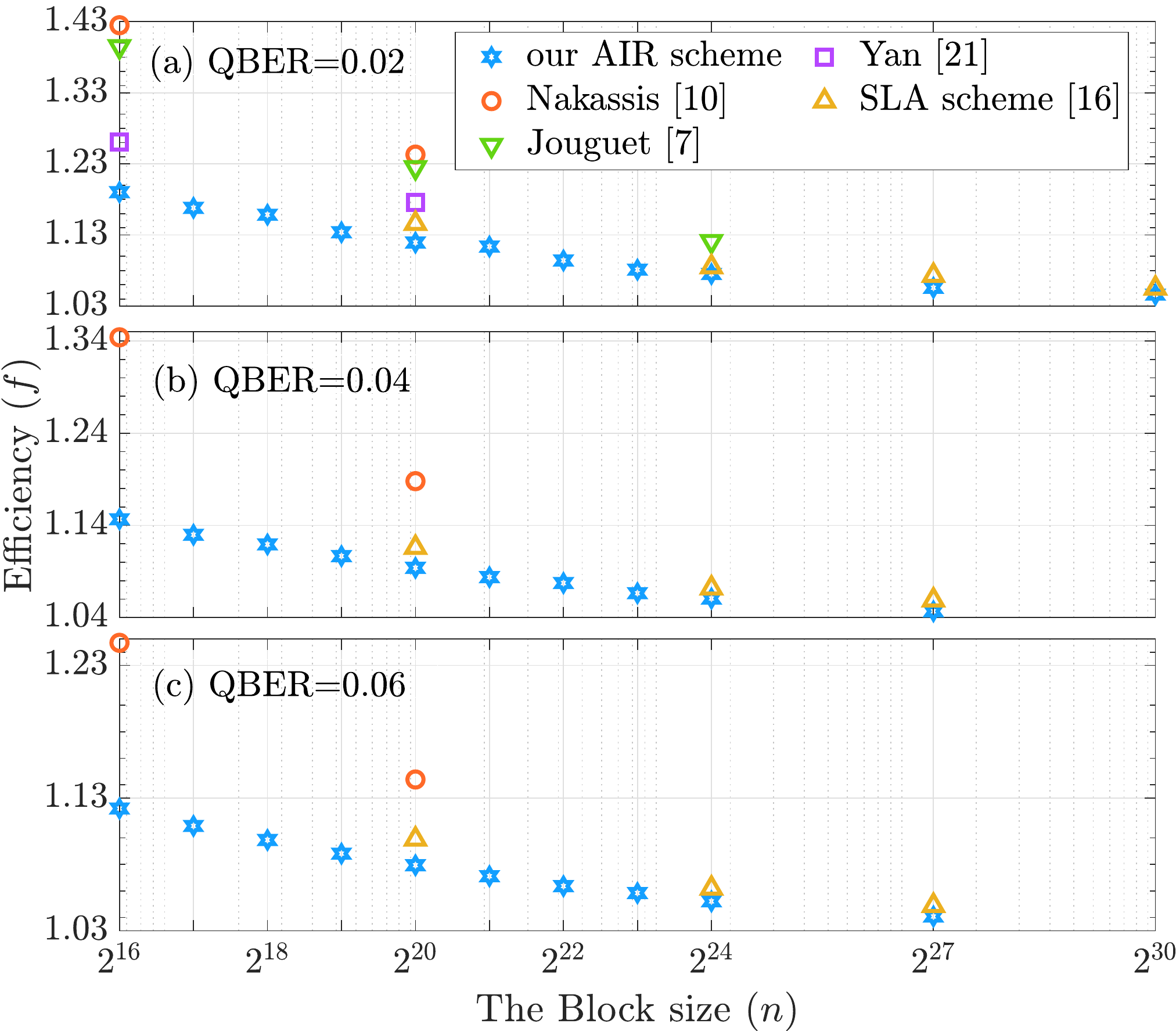}
	\caption{The efficiency of the AIR scheme, compared with the previous polar codes-based IR schemes. Here $n$ is varying from $2^{16}$ to $2^{30}$, $R_m = 4$ and $E_\mu =0.02, 0.04, 0.06$.}
	\label{fig.qber_all_16}
\end{figure}

The efficiency of the proposed AIR scheme, compared with the previous implemented polar codes-based IR schemes~\cite{RN2,RN12,RN25,RN65}, are shown in FIG.~\ref{fig.qber_all_16}, where $n$ is varying from $2^{16}$ to $2^{30}$, $R_m = 4$ and $E_\mu =0.02, 0.04, 0.06$. Meanwhile, the failure probability of the proposed AIR scheme is around $10^{-8}$ while the majority of the previous schemes stay at the level of $10^{-3}$.

As shown in FIG.~\ref{fig.qber_all_16}, the efficiency of our proposed AIR scheme is further closer to the Shannon limit than the state-of-the-art scheme (SLA), especially when performed with smaller block sizes~\cite{RN2}. The efficiency, achieved by the AIR scheme with $n=$\SI{256}{Kb}, is comparable to the efficiency of the previous SLA scheme, which has to be performed with 4 times larger block size. Moreover, the efficiency of the proposed AIR scheme reaches 1.046, when the block size is \SI{1}{Gb} and the quantum bit error rate of 0.02. More detailed efficiency analysis results are shown in Appendix \ref{sec:appA}.

\subsection{Secure key rate estimation with the AIR scheme}
\label{section:simulate R}

The secure key rates of the practical QKD systems can be improved with higher information reconciliation efficiency. We estimate the secure key rate of QKD system performed with the AIR scheme with the continuous-wave (CW) BBM92 protocol~\cite{RN78}. The estimation parameters are given in Table~\ref{tab.CW_simulation}.

\begin{table}[htbp]
  \begin{center}
    \renewcommand{\arraystretch}{1.2} %
    \caption{The parameters for secure key rate estimation with the continuous-wave BBM92 protocol. $e^{\rm{pol}}$ is the individual polarization error probability, ${\rm{DCR}}$ is the dark count rate of each side, $t_{\rm{cc}}$ is the coincidence window, and $t_{\rm{\Delta}}$ is the full width at half maximum of the correlation histogram. $f_{\rm{AIR}}$ is the efficiency of the AIR scheme and $f_{\rm{c}}$ is the efficiency commonly used previously. The brightness of the entangled photon source is optimized for each channel loss.} 
    \label{tab.CW_simulation} 
    \setlength{\tabcolsep}{5pt}
    {
    \begin{tabular}{c c c c c c} 
      \toprule[1pt]
      $e^{\rm{pol}}$ &${\rm{DCR}}$ &$t_{\rm{cc}}$ &$t_{\rm{\Delta}}$ & $f_{\rm{AIR}}$ &$f_{\rm{c}}$ \\
      \hline
      2\% &\SI{200}{cps} & $\SI{140}{ps}$ &\SI{140}{ps} &1.046  &1.200 \\
      \hline
    \end{tabular}}
  \end{center}
\end{table}

\begin{figure}[!ht]
	\centering
	\includegraphics[width=0.48\textwidth]{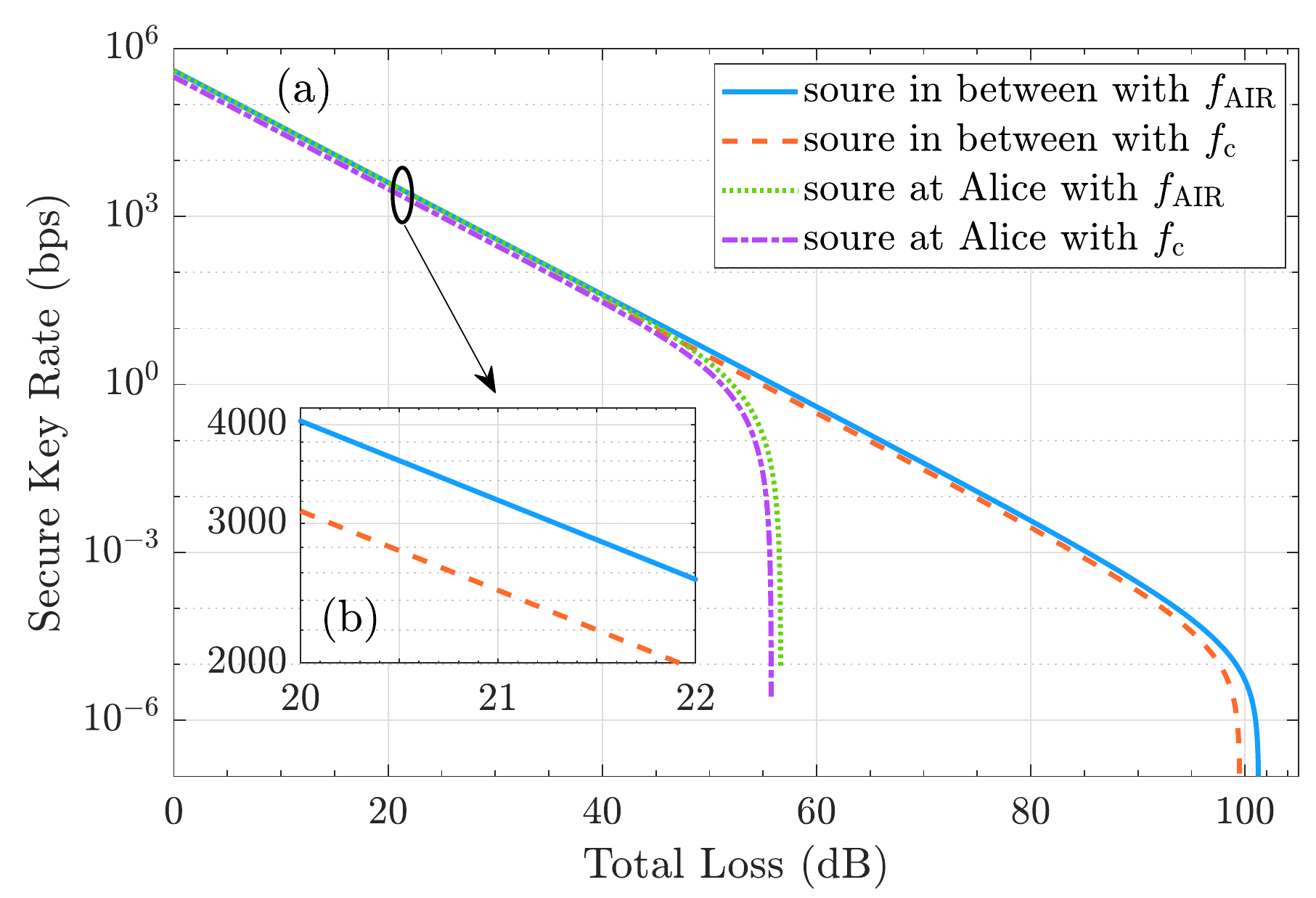}\caption{(a) The secure key rate with the efficiency of 1.046 for the AIR scheme based QKD system and 1.2 for the previously practical systems. (b) Zoomed in figure of the source in between with total loss varying from 20.0 to 22.0 dB.}
	\label{fig.compare 1.046 and 1.2}
\end{figure}

For the estimation, the efficiency of the AIR scheme is set to 1.046 and the commonly used value for previously practical systems is 1.2~\cite{RN89, RN117}. As shown in FIG.~\ref{fig.compare 1.046 and 1.2}, the secure key rate of QKD systems performed with the AIR scheme can be increased at least 30\% than the previous QKD systems with $f=1.2$, which further eradicates the performance gap between theory and implementation for QKD systems.

\section{Conclusion}
In this article, we propose an appending information reconciliation (AIR) scheme for quantum key distribution (QKD) system, which mainly includes initialization phase and appending reconciliation phase. In the initialization phase, Alice constructs the codeword and sends the cyclic redundancy check (CRC) tag of the codeword to Bob. In the appending reconciliation phase, according to the optimized and pre-shared frozen vectors, Alice appends the syndrome information, which is combined with the bit values indicated by the frozen vectors, to Bob until the CRC values of the codewords from Alice and Bob are equal, or the interactive rounds reach the maximum round number. 

The experimental results show that compared with the previous polar codes-based IR schemes, the proposed AIR scheme reaches high efficiency and ultra-low failure probability simultaneously with the same block size and QBER. When the efficiency of the AIR scheme is comparable to the efficiency of the state-of-the-art scheme, the AIR scheme only needs to be performed with 4 times smaller block size. Therefore, our AIR scheme is more executable in the practical QKD system with the finite block size. In particular, the AIR scheme can achieve the efficiency of $1.046$, when the block size is \SI{1}{Gb} and quantum bit error rate of 0.02, which is closer to the Shannon limit.

\textbf{\textit{Acknowledgements}} This work was supported by the National Natural Science Foundation of China under Grant No. 61972410, the Research Plan of National University of Defense Technology under Grant No. ZK19-13 and No. 19-QNCXJ-107 and the Postgraduate Scientific Research Innovation Project of Hunan Province under Grant No. CX20200003.

\textbf{\textit{Author contributions}}
HZ and BYT contributed equally to this paper. HZ, BYT and BL proposed the scheme and wrote the paper. WRY, HZ and HC designed the experiments. HC, HZ and HCY performed the experiments, and BL, BYT, SCL participated in the data analysis. This work was supervised by WRY and BL. All authors discussed and revised the manuscript.

\bibliographystyle{plainnat}

\begin{thebibliography}{23}
	\providecommand{\natexlab}[1]{#1}
	\providecommand{\url}[1]{\texttt{#1}}
	\expandafter\ifx\csname urlstyle\endcsname\relax
	\providecommand{\doi}[1]{doi: #1}\else
	\providecommand{\doi}{doi: \begingroup \urlstyle{rm}\Url}\fi
	
	\bibitem{RN4}
	Erdal Arikan.
	\newblock Channel polarization: A method for constructing capacity-achieving
	codes for symmetric binary-input memoryless channels.
	\newblock \emph{IEEE Transactions on Information Theory}, 55\penalty0
	(7):\penalty0 3051--3073, 2009.
	\newblock \doi{10.1109/TIT.2009.2021379}.
	
	\bibitem{RN70}
	Charles~H. Bennett, François Bessette, Gilles Brassard, Louis Salvail, and
	John Smolin.
	\newblock Experimental quantum cryptography.
	\newblock \emph{Journal of Cryptology}, 5\penalty0 (1):\penalty0 3--28, 1992.
	\newblock ISSN 1432-1378.
	\newblock \doi{10.1007/BF00191318}.
	
	\bibitem{RN27}
	Gilles Brassard and Louis Salvail.
	\newblock Secret-key reconciliation by public discussion.
	\newblock In Tor Helleseth, editor, \emph{Advances in Cryptology — EUROCRYPT
		’93}, pages 410--423. Springer Berlin Heidelberg, 1994.
	\newblock ISBN 978-3-540-48285-7.
	\newblock \doi{10.1007/3-540-48285-7_35}.
	
	\bibitem{RN7}
	Y.~A. Chen, A.~N. Zhang, Z.~Zhao, X.~Q. Zhou, C.~Y. Lu, C.~Z. Peng, T.~Yang,
	and J.~W. Pan.
	\newblock Experimental quantum secret sharing and third-man quantum
	cryptography.
	\newblock \emph{Phys Rev Lett}, 95\penalty0 (20):\penalty0 200502, 2005.
	\newblock ISSN 0031-9007 (Print) 0031-9007 (Linking).
	\newblock \doi{10.1103/PhysRevLett.95.200502}.
	
	\bibitem{RN14}
	D.~Elkouss, A.~Leverrier, R.~Alleaume, and J.~J. Boutros.
	\newblock Efficient reconciliation protocol for discrete-variable quantum key
	distribution.
	\newblock In \emph{2009 IEEE International Symposium on Information Theory},
	pages 1879--1883, 2009.
	\newblock ISBN 2157-8117.
	\newblock \doi{10.1109/ISIT.2009.5205475}.
	
	\bibitem{RN153}
	Christopher Huth, René Guillaume, Thomas Strohm, Paul Duplys, Irin~Ann Samuel,
	and Tim Güneysu.
	\newblock Information reconciliation schemes in physical-layer security: A
	survey.
	\newblock \emph{Computer Networks}, 109:\penalty0 84--104, 2016.
	\newblock ISSN 1389-1286.
	\newblock \doi{10.1016/j.comnet.2016.06.014}.
	
	\bibitem{RN12}
	P.~Jouguet and S.~Kunz-Jacques.
	\newblock High performance error correction for quantum key distribution using
	polar codes.
	\newblock \emph{Quantum Information \& Computation}, 14\penalty0
	(3-4):\penalty0 329--338, 2014.
	\newblock ISSN 1533-7146.
	\newblock \doi{10.48550/ARXIV.1204.5882}.
	
	\bibitem{RN18}
	H.~K. Lo, X.~Ma, and K.~Chen.
	\newblock Decoy state quantum key distribution.
	\newblock \emph{Phys Rev Lett}, 94\penalty0 (23):\penalty0 230504, 2005.
	\newblock ISSN 0031-9007 (Print) 0031-9007 (Linking).
	\newblock \doi{10.1103/PhysRevLett.94.230504}.
	
	\bibitem{RN1}
	J.~Martinez-Mateo, D.~Elkouss, and V.~Martin.
	\newblock Key reconciliation for high performance quantum key distribution.
	\newblock \emph{Sci Rep}, 3:\penalty0 1576, 2013.
	\newblock ISSN 2045-2322 (Electronic) 2045-2322 (Linking).
	\newblock \doi{10.1038/srep01576}.
	
	\bibitem{RN65}
	Anastase Nakassis and Alan Mink.
	\newblock Polar codes in a qkd environment.
	\newblock \emph{Quantum Information and Computation XII}, 9123:\penalty0
	912305, 2014.
	\newblock \doi{10.1117/12.2050919}.
	
	\bibitem{RN78}
	Sebastian~Philipp Neumann, Thomas Scheidl, Mirela Selimovic, Matej Pivoluska,
	Bo~Liu, Martin Bohmann, and Rupert Ursin.
	\newblock Model for optimizing quantum key distribution with continuous-wave
	pumped entangled-photon sources.
	\newblock \emph{Physical Review A}, 104\penalty0 (2):\penalty0 022406, 2021.
	\newblock \doi{10.1103/PhysRevA.104.022406}.
	
	\bibitem{RN99}
	David Pearson.
	\newblock High-speed qkd reconciliation using forward error correction.
	\newblock \emph{AIP Conference Proceedings}, 734:\penalty0 299--302, 11 2004.
	\newblock \doi{10.1063/1.1834439}.
	
	\bibitem{RN97}
	Li~Qiong, Le~Dan, Mao Haokun, Niu Xiamu, Liu Tian, and Guo Hong.
	\newblock Study on error reconciliation in quantum key distribution.
	\newblock \emph{Quantum Information and Computation}, 14:\penalty0 1117--1135,
	2014.
	\newblock \doi{10.5555/2685164.2685169}.
	
	\bibitem{RN96}
	RENATO RENNER.
	\newblock Security of quantum key distribution.
	\newblock \emph{International Journal of Quantum Information}, 06\penalty0
	(01):\penalty0 1--127, 2008.
	\newblock \doi{10.1142/s0219749908003256}.
	
	\bibitem{RN37}
	I.~Tal and A.~Vardy.
	\newblock How to construct polar codes.
	\newblock \emph{Ieee Transactions on Information Theory}, 59\penalty0
	(10):\penalty0 6562--6582, 2013.
	\newblock ISSN 0018-9448.
	\newblock \doi{10.1109/Tit.2013.2272694}.
	
	\bibitem{RN2}
	Bang-Ying Tang, Bo~Liu, Wan-Rong Yu, and Chun-Qing Wu.
	\newblock Shannon-limit approached information reconciliation for quantum key
	distribution.
	\newblock \emph{Quantum Information Processing}, 20\penalty0 (3), 2021.
	\newblock ISSN 1570-0755 1573-1332.
	\newblock \doi{10.1007/s11128-020-02919-8}.
	
	\bibitem{RN60}
	Metin Toyran, Mustafa Toyran, and S~{\"O}zt{\"u}rk.
	\newblock New approaches to increase efficiency of cascade information
	reconciliation protocol.
	\newblock In \emph{7th International Conference on Quantum Cryptography,
		Cambridge, UK}, 2017.
	
	\bibitem{RN94}
	X.~Wang, Y.~Zhang, S.~Yu, and H.~Guo.
	\newblock High speed error correction for continuous-variable quantum key
	distribution with multi-edge type ldpc code.
	\newblock \emph{Sci Rep}, 8\penalty0 (1):\penalty0 10543, 2018.
	\newblock ISSN 2045-2322.
	\newblock \doi{10.1038/s41598-018-28703-4}.
	
	\bibitem{RN89}
	Sören Wengerowsky, Siddarth~Koduru Joshi, Fabian Steinlechner, Julien~R.
	Zichi, Sergiy~M. Dobrovolskiy, René van~der Molen, Johannes W.~N. Los, Val
	Zwiller, Marijn A.~M. Versteegh, Alberto Mura, Davide Calonico, Massimo
	Inguscio, Hannes Hübel, Liu Bo, Thomas Scheidl, Anton Zeilinger, André
	Xuereb, and Rupert Ursin.
	\newblock Entanglement distribution over a 96-km-long submarine optical fiber.
	\newblock \emph{Proceedings of the National Academy of Sciences}, 116\penalty0
	(14):\penalty0 6684--6688, 2019.
	\newblock \doi{10.1073/pnas.1818752116}.
	
	\bibitem{RN23}
	H.~Yan, T.~N. Ren, X.~Peng, X.~X. Lin, W.~Jiang, T.~Liu, and H.~Guo.
	\newblock Information reconciliation protocol in quantum key distribution
	system.
	\newblock \emph{Icnc 2008: Fourth International Conference on Natural
		Computation, Vol 3, Proceedings}, pages 637--641, 2008.
	\newblock \doi{10.1109/Icnc.2008.755}.
	
	\bibitem{RN25}
	S.~L. Yan, J.~D. Wang, J.~B. Fang, L.~Jiang, and X.~Wang.
	\newblock An improved polar codes-based key reconciliation for practical
	quantum key distribution.
	\newblock \emph{Chinese Journal of Electronics}, 27\penalty0 (2):\penalty0
	250--255, 2018.
	\newblock ISSN 1022-4653.
	\newblock \doi{10.1049/cje.2017.07.006}.
	
	\bibitem{RN30}
	Li~Yang.
	\newblock One-way information reconciliation schemes of quantum key
	distribution.
	\newblock \emph{Cybersecurity}, 2\penalty0 (1), 2019.
	\newblock ISSN 2096-4862.
	\newblock \doi{10.1186/s42400-019-0033-z}.
	
	\bibitem{RN117}
	Juan Yin, Yuan Cao, Yu-Huai Li, Ji-Gang Ren, Sheng-Kai Liao, Liang Zhang,
	Wen-Qi Cai, Wei-Yue Liu, Bo~Li, Hui Dai, Ming Li, Yong-Mei Huang, Lei Deng,
	Li~Li, Qiang Zhang, Nai-Le Liu, Yu-Ao Chen, Chao-Yang Lu, Rong Shu, Cheng-Zhi
	Peng, Jian-Yu Wang, and Jian-Wei Pan.
	\newblock Satellite-to-ground entanglement-based quantum key distribution.
	\newblock \emph{Physical Review Letters}, 119\penalty0 (20):\penalty0 200501,
	2017.
	\newblock \doi{10.1103/PhysRevLett.119.200501}.
	
\end{thebibliography}

\onecolumn\newpage
\appendix

\section{The detailed estimation results}
\label{sec:appA}
We test the block size ranging from $2^{16}$ to $2^{27}$, with the SCL decoder and QBER ranging from 0.01 to 0.12 with an interval of 0.01, while the length of CRC is set as 64, the list size of the SCL decoder is set as 16, the decoding failure probability in the $R_m$-th round $\varepsilon_{R_m}$ is $10^{-8}$ and the maximum round number $R_m$ is 4. The average execution rounds shows in Table.~\ref{tab.polar_round_av} and the reconciliation efficiencies show in Table.~\ref{tab.polar_all}.

\begin{table*}[!hbp]
  \begin{center}
    \setlength{\tabcolsep}{3.8pt}%
    \renewcommand{\arraystretch}{1.5} %
    \caption{The average execution rounds of AIR scheme.} 
    \label{tab.polar_round_av}
    \begin{tabular}{c l l l l l l l l l l l l} 
      \toprule[1pt]
      $n$ &0.01 &0.02 &0.03 &0.04 &0.05 &0.06 &0.07 &0.08 &0.09 &0.10 &0.11 &0.12\\
      \hline
      $2^{16}$ &1.635 &1.677 &1.639 &1.612 &1.758 &1.643 &1.650 &1.651 &1.647 &1.693 &1.772 &1.775\\
      \hline      
      $2^{17}$ &1.720 &1.613 &1.600 &1.493 &1.735 &1.550 &1.618 &1.808 &1.468 &1.760 &1.685 &1.500 \\
      \hline
      $2^{18}$ &1.720 &1.811 &1.658 &1.623 &1.654 &1.536 &1.673 &1.593 &1.597 &1.865 &1.665 &1.600 \\
      \hline
      $2^{19}$ &1.818 &1.644 &1.653 &1.393 &1.458 &1.639 &1.666 &1.634 &1.639 &1.573 &1.510 &1.868 \\
      \hline
      $2^{20}$ &1.632 &1.637 &1.633 &1.518 &1.652 &1.774 &1.440 &1.576 &1.732 &1.434 &1.639 &1.607 \\
      \hline
      $2^{21}$ &1.761 &1.509 &1.549 &1.551 &1.686 &1.435 &1.615 &1.532 &1.668 &1.668 &1.702 &1.453 \\
      \hline
      $2^{22}$ &1.657 &1.640 &1.544 &1.784 &1.597 &1.581 &1.624 &1.634 &1.579 &1.589 &1.475 &1.492 \\
      \hline
      $2^{23}$ &1.811 &1.372 &1.676 &1.542 &1.650 &1.702 &1.596 &1.546 &1.695 &1.669 &1.732 &1.435 \\
      \hline
      $2^{24}$ &1.515 &1.363 &1.547 &1.594 &1.770 &1.680 &1.494 &1.815 &1.499 &1.708 &1.637 &1.357 \\
      \hline
      $2^{27}$ &1.660 &1.380 &1.550 &1.690 &1.470 &1.300 &1.700 &1.650 &1.890 &1.630 &1.790 &1.920 \\
      \hline
    \end{tabular}
  \end{center}
\end{table*}

\begin{table*}[!hbp]
	\begin{center}
		\setlength{\tabcolsep}{3.8pt}%
		\renewcommand{\arraystretch}{1.5} %
		\caption{The reconciliation efficiency of AIR scheme.} 
		\label{tab.polar_all}
		\begin{tabular}{c l l l l l l l l l l l l} 
			\toprule[1pt]
			$n$ &0.01 &0.02 &0.03 &0.04 &0.05 &0.06 &0.07 &0.08 &0.09 &0.10 &0.11 &0.12\\
			\hline
			$2^{16}$ &1.241 	&1.190 	&1.171 	&1.147 	&1.142 	&1.122 	&1.114 	&1.104 	&1.097 	&1.105 	&1.089 	&1.079 \\
			\hline
			$2^{17}$ &1.206	&1.168	&1.147	&1.130	&1.118	&1.109	&1.099	&1.092	&1.087	&1.082	&1.075	&1.072\\
			\hline
			$2^{18}$ &1.189	&1.158	&1.132	&1.119	&1.107	&1.098	&1.091	&1.085	&1.079	&1.074	&1.070	&1.065\\
			\hline
			$2^{19}$  &1.167  &1.134  &1.117  &1.107  &1.095  &1.088  &1.082  &1.076  &1.071  &1.067  &1.062  &1.059\\
			\hline
			$2^{20}$  &1.147 &1.120  &1.105  &1.094  &1.085  &1.079  &1.073  &1.068 &1.064  &1.060  &1.056  &1.053\\
			\hline
			$2^{21}$  &1.133  &1.114  &1.094 &1.083 &1.078  &1.071  &1.066  &1.062  &1.059  &1.054  &1.051  &1.048\\
			\hline
			$2^{22}$  &1.117  &1.094  &1.083  &1.078  &1.070  &1.064  &1.060  &1.059  &1.053  &1.051  &1.047  &1.044\\
			\hline
			$2^{23}$  &1.105  &1.081  &1.074  &1.067  &1.064  &1.058  &1.054  &1.051 &1.048  &1.045  &1.043  &1.040\\
			\hline
			$2^{24}$  &1.090  &1.075  &1.067  &1.061  &1.056  &1.052  &1.049  &1.049  &1.044  &1.041  &1.039  &1.037\\
			\hline
			$2^{27}$  &1.070  &1.056  &1.051  &1.047 &1.045  &1.040  &1.039  &1.037  &1.036  &1.034  &1.033  &1.031\\
			\hline
		\end{tabular}
	\end{center}
\end{table*}

\section{Detailed analysis of probabilities}

\subsection{The decoding success probability}
\label{sectionAP:success probability}

The decoding success probability in the $i$-th round, which means the decoding procedure fails with the $S_D^{i-1}$ and succeeds with the $S_D^{i}$, can be calculated as

\begin{equation}
\begin{aligned}
\mathrm{Pr}\left[(U^\prime = U | S_D^i) \cap (U^\prime \neq U | S_D^{i-1})\right] 
&=\mathrm{Pr}\left[(U^\prime = U | S_D^i) \cap \left((U^\prime |S_D^{i-1}) - (U^\prime = U | S_D^{i-1})\right)\right]\\
&=\mathrm{Pr}\left[(U^\prime = U | S_D^i) \cap (U^\prime |S_D^{i-1}) - (U^\prime = U | S_D^i) \cap (U^\prime = U | S_D^{i-1})\right]\\
&=\mathrm{Pr}\left[(U^\prime = U | S_D^i) - (U^\prime = U | S_D^{i-1})\right]\\
&=\mathrm{Pr}(U^\prime = U | S_D^i) - \mathrm{Pr}(U^\prime = U | S_D^{i-1})\\
&= (1-\varepsilon_{i}) - (1-\varepsilon_{i-1})\\
&= \varepsilon_{i-1} - \varepsilon_i
\end{aligned}.
\end{equation}

\subsection{The overall failure probability}
\label{sectionAP:pf}

\textbf{Case \Rmnum{1}.} $\mathrm{CRC}(U^{\prime})=T$ but $U^\prime \neq U$ and $i\leq R_m$, which means the IR procedure fails the CRC check in the previous $(i-1)$ rounds and succeeds in the $i$-th round, but the decoding procedure fails. The failure probability $\varepsilon_\mathrm{I}$ of this case is calculated as

\begin{equation}
\begin{aligned}
  \varepsilon_\mathrm{I} &= \sum_{i=1}^{R_m} \mathrm{Pr}\left\{\left[\mathrm{CRC}(U^\prime) \neq T|\left(U^\prime \neq U,(1,2,\cdots,i-1)\right)\right] \cap \left[\mathrm{CRC}(U^\prime)=T|(U^\prime \neq U,i)\right]\right\} \\ 
  &=\sum_{i=1}^{R_m} \mathrm{Pr}\left[\mathrm{CRC}(U^\prime) \neq T|(1,2,\cdots,i-1)\right] \cdot \mathrm{Pr}\left[\mathrm{CRC}(U^\prime)=T|(U^\prime \neq U,i)\right] \\
  &= \sum_{i=1}^{R_m} (1-\varepsilon_\mathrm{CRC})^{i-1}\cdot{\varepsilon_i}\varepsilon_\mathrm{CRC}\\
  &=\varepsilon_\mathrm{CRC}\sum_{i=1}^{R_m}{\varepsilon_i(1-\varepsilon_\mathrm{CRC})^{i-1}}\\
\end{aligned}.  
\end{equation}

\textbf{Case \Rmnum{2}.} $\mathrm{CRC(U^{\prime})}\neq T$ and $i = R_m$, which means the IR procedure fails the CRC check in the previous ($i-1$) rounds and aborts in the maximum round $R_m$. In this case, the failure probability $\varepsilon_\mathrm{II}$ can be calculated as
\begin{equation}
  \begin{aligned}
  \varepsilon_\mathrm{II} &= \mathrm{Pr}\left\{\left[\mathrm{CRC}(U^\prime) \neq T|\left(U^\prime \neq U,(1,2,\cdots,i-1)\right)\right] \cap \left[\mathrm{CRC}(U^\prime)\neq T|(U^\prime \neq U,i)\right]\right\}\\ 
  &=\mathrm{Pr}\left[\mathrm{CRC}(U^\prime) \neq T|(1,2,\cdots,i-1)\right] \cdot  \mathrm{Pr}\left[\mathrm{CRC}(U^\prime)\neq T|(U^\prime \neq U,i)\right] \\
  &=(1-\varepsilon_\mathrm{CRC})^{i-1}\cdot\varepsilon_{i}(1-\varepsilon_\mathrm{CRC})\\
  &=\varepsilon_{i}(1-\varepsilon_\mathrm{CRC})^{i}
  \end{aligned}.  
\end{equation}

\subsection{The probability which the IR procedure stops in each round}\label{sectionAP:round_av}

\textbf{Case \Rmnum{1}.} $i<R_m$, which means the IR procedure is failed the CRC check in the previous $(i-1)$ rounds and succeeds in the $i$-th round. The $P_s^i$ can be calculated as
\begin{equation}
  \begin{aligned}
  P_s^i
  &= \mathrm{Pr}\left\{\left[\mathrm{CRC}(U^\prime) \neq T|\left(U^\prime \neq U,(1,2,\cdots,i-1)\right)\right] \cap \left[U^\prime = U,i\right]\right\}\\ 
  &+ \mathrm{Pr}\left\{\left[\mathrm{CRC}(U^\prime) \neq T|\left(U^\prime \neq U,(1,2,\cdots,i-1)\right)\right] \cap\left[\mathrm{CRC}(U^\prime)=\mathrm{CRC}(U)|(U^\prime \neq U,i)\right]\right\}\\ 
  &=\mathrm{Pr}\left[\mathrm{CRC}(U^\prime) \neq T|(1,2,\cdots,i-1)\right] \cdot\mathrm{Pr}\left[U^\prime = U,i\right]\\ 
  &+ \mathrm{Pr}\left[\mathrm{CRC}(U^\prime) \neq T|(1,2,\cdots,i-1)\right]\cdot\mathrm{Pr}\left[\mathrm{CRC}(U^\prime)=\mathrm{CRC}(U)|(U^\prime \neq U,i)\right]\\
  &= (1-\varepsilon_{\mathrm{CRC}})^{i-1}\cdot\left[(\varepsilon_{i-1} -\varepsilon_{i})\right] \\
  &+(1-\varepsilon_{\mathrm{CRC}})^{i-1}\cdot\varepsilon_{i}\varepsilon_{\mathrm{CRC}}\\
  &=\varepsilon_{i-1}(1-\varepsilon_{\mathrm{CRC}})^{i-1} - \varepsilon_ {i}(1-\varepsilon_{\mathrm{CRC}})^{i}   
  \end{aligned}.
\end{equation}

\textbf{Case \Rmnum{2}.} $i=R_m$, which means the IR procedure is stopped in the the $R_m$-th round, and the IR procedure fails the CRC check in the previous ($i-1$) rounds, the $P_s^i$ can be calculated as
\begin{equation}
  \begin{aligned}
    P_s^i &=  \mathrm{Pr}\left[\mathrm{CRC}(U^\prime) \neq T|\left(U^\prime \neq U,(1,2,\cdots,i-1)\right)\right]\\
    &=\varepsilon_{i-1}(1-\varepsilon_{\mathrm{CRC}})^{i-1}\\
  \end{aligned}.
\end{equation}

\end{document}